\documentclass[10pts,showpacs,preprint,aps]{revtex4}
\usepackage{graphicx}% Include figure files
\usepackage{dcolumn}% Align table columns on decimal point
\usepackage{bm}% bold math
\usepackage{amssymb}
\usepackage{amsmath}
\begin{document}
\setcounter{page}{1}
\vskip 2cm
\title 
{The Astrophysical Scales Set by the Cosmological Constant, Black-Hole Thermodynamics and Non-Linear Massive Gravity}
\author
{Ivan Arraut}
\affiliation{$^1$Department of Physics, Faculty of Science, Tokyo University of Science,
1-3, Kagurazaka, Shinjuku-ku, Tokyo 162-8601, Japan}
\affiliation{$^2$Department of Physics, Osaka University, Toyonaka, Osaka 560-0043, Japan}
\affiliation{$^3$Theory Center, Institute of Particle and Nuclear Studies, KEK Tsukuba, Ibaraki, 305-0801, Japan}

\begin{abstract}
We calculate explicitly the black-hole temperature for the Schwarzschild de-Sitter solution inside massive gravity by defining the Killing-vector in the direction of the St\"uckelberg function. \mbox{We then} consider the conditions which an observer in massive gravity has to obey in order to agree with the standard results of General Relativity.  
%The positive cosmological constant sets scales which in combination with other parameters of the theory, can be of astrophysical order of magnitude. Here we analyze the scale where the gravitational effects of matter (including dynamics) are equivalent to the repulsive ones due to the cosmological constant. We then compare such scale with the role of the Vainshtein radius inside the non-linear massive gravity theory formulations.   
\end{abstract}
\maketitle 
\section{Introduction}
The analytic extension for the Schwarzschild de-Sitter (S-dS) space has been done in \cite{1}. In such a case, the scale $r_0=\left(\frac{3}{2}r_s r_\Lambda^2\right)^{1/3}$, with $r_\Lambda=1/\sqrt{\Lambda}$ and $r_s=2GM$ defining %is GM an unit or not? A: It is a scale.  should it be italic or not? A. It should be inside the symbol $$. A. Is there space between 2 and GM? There is no space between 2 and GM.
the cosmological constant scale and the Schwarzschild scale respectively, appeared as the distance where the $0-0$ %please check whether it is endash between them.A. It is correct in this way.
component of the S-dS metric takes a minimum value. Some interesting analysis about the physical consequences of this scale, has been done in \cite{2, 3, 3-2, 4, 5, 6, 7}. In massive gravity theories, a similar scale appears and it is called Vainshtein radius \cite{Vainshtein, Vainshtein-2}. In General Relativity (GR), the scale $r_0$ represents the location of the static observer. This is the case because in the S-dS space we cannot define as static observers, those located at the infinity as we did for the cases of asymptotically flat spaces. The only observer who does not feel gravity (excluding free-falling condition)  and as a consequence, the only observer who can be considered as static, is the one located at $r_0$. This was the key point for the analysis done by Bousso and Hawking in order to find the appropriate expression for the temperature of a black hole immersed inside a de-Sitter space \cite{3}. The location of the static observers at the scale $r_0$ is equivalent to a normalization of the time-like Killing vector with respect to such observers. This is the main reason because of which in \cite{3, 4} there is a minimal temperature for the black hole in the S-dS space. \mbox{Note that} since the scale $r_0$ represents a local maximum if we find the effective potential for a test particle moving around the source \cite{7, 8, 9, 10, 11, 11-2, 11-3, 11-4, 11-5, 11-6, 12, 12-2}, then it represents the distance after which there are no bound orbits. This~means that the cosmological constant in GR is relevant at scales larger than $r_0$, \mbox{this is} just analogous to the role of the Vainshtein radius ($r_V$) in massive gravity, where $r_V$ is the scale after which the effect of the extra-degrees of freedom becomes to be relevant. Here we analyze the S-dS solution in massive gravity and we find the conditions under which the black-hole temperature, as it is defined by static observers in massive gravity, agrees with the black-hole temperature as it is measured by analogous observers in the standard formulation of gravity (GR). Since massive gravity is supposed to approach to GR at scales where the gravitational field is strong, then it is expected that the same amount of particles at the event horizon are created in massive gravity if we compare with the case formulated in GR. However, this does not guarantee that the observers defined in both theories and satisfying the same conditions, will agree with the final result. This is the case because in massive gravity the St\"uckelberg function $T_0(r,t)$ defines a preferred notion of time. This~is equivalent to say that in this theory the extra-degrees of freedom will create a distortion in the notions of time \cite{NR, NR-2}, \mbox{and as a consequence}, an apparent non-conservation of the energy-momentum tensor \cite{Komar}. \mbox{Then in general,} observers defined in massive gravity will measure different temperatures with respect to the observers living in the theory of GR, even if they are defined by using the same conditions. However, we can find the conditions under which the observers defined in GR will agree with those defined in massive gravity. For this purpose, it is convenient to define in massive gravity the black-hole temperature by using the Killing vector in the direction of the St\"uckelberg function $T_0(r,t)$. Only the observers defining the time coordinate as $T_0(r,t)$ will agree in the results of temperature obtained in the GR case. Any other observer, defining the time arbitrarily, will disagree with the results obtained by equivalent observers in GR. Then the conditions under which the temperature as it is defined in massive gravity agrees with the one defined in GR, impose some constraints in the functional behavior of the St\"uckelberg fields. The paper is organized as follows: In~Section~\ref{eq:Lag MOND}, we~derive the standard solution of S-dS inside GR together with the conserved quantities for test particles and the event horizons for the solution. In~Section~\ref{eq:Critical}, we find the critical points for the effective potential of a test particle inside GR. The critical points correspond to the circular orbit conditions obtained from the derivatives of the effective potential. In the Section~\ref{eq:Momoko Cheung}, we explain the expression for the black-hole temperature inside GR by using the appropriate normalization for the time-like Killing vector. This~normalization imposes a limit for the minimum value which the black-hole temperature can take. In~Section~\ref{eq:Final1}, we explain the S-dS solution in the non-linear massive gravity theory and then we define the conserved quantities and equations of motion for a test particle. In Section~\ref{BHTINLMG}, we derive the black-hole temperature in massive gravity and then we find the necessary conditions which the St\"uckelberg functions, as they are defined by the static observers, have to satisfy such that we get the standard results of GR corresponding to the black-hole temperature inside massive gravity. In massive gravity we use the same normalization of the Killing vector as in GR, but this time we define the Killing vector in the direction of the St\"uckelberg function instead of defining it in the time-direction. {In Section~\ref{Finalsection}, we make the expansion up to second order of the action in massive gravity in a free-falling frame of reference. This corresponds to the simplest case and it provides the scenario for understanding how the number of degrees of freedom can change for different observers and how the masses corresponding to different modes fluctuates}. Finally, in Section~\ref{eq:conclusions}, we conclude.             
      
\section{The Schwarzschild De-Sitter Metric in Static Coordinates}   \label{eq:Lag MOND}

The Schwarzschild-de Sitter (S-dS) metric in static coordinates, is defined by
\begin{equation}   \label{eq:Sdsm}
ds^2=-e^{\nu(r)}dt^2+e^{-\nu(r)}dr^2+r^2d\theta^2+r^2\sin^2\theta d\phi^2,
\end{equation}
where
\begin{equation}   \label{eq:e}
e^{\nu(r)}=1-\frac{r_s}{r}-\frac{r^2}{3r_\varLambda^2}.
\end{equation}

Here $r_s=2GM$ %is it an unit or not? should it be italic or not? Is there space between 2 and GM? A. I have corrected it.
 is the Schwarzschild radius and $r_\Lambda=\frac{1}{\sqrt{\Lambda}}$ is the cosmological constant scale. If we want to find the equations of motion for the S-dS metric, the work is simplified if we use the symmetries of the solution. We know that there are four killing vectors, three for spatial symmetry and one for time translations. Each of these Killing vectors will be related to a constant of motion for a~free particle \cite{13}. If $K^\mu$ is a Killing vector, it is well known that
\begin{equation}   \label{eq:CQa}
K_\mu \frac{dx^\mu}{d\lambda}=Constant,
\end{equation}
is a constant of motion. We can define an additional constant of motion given by 
\begin{equation}   \label{eq:KE}
g_{\mu \nu}\frac{dx^\mu}{d\lambda}\frac{dx^\nu}{d\lambda}=-\epsilon.
\end{equation}

This is a constant of motion along the path of the test particle \cite{13}. In Equation~(\ref{eq:KE}) $\epsilon=1$ for massive particles and $\epsilon=0$ for massless particles. The quantity associated with the invariance under spatial rotations is the angular momentum. We can think about the angular momentum as a~three-vector with magnitude (one component) and direction (two components). Conservation of the direction of angular momentum means that the particle will move over a plane, we can choose this to be the equatorial plane of our coordinate system. If the particle is not initially over the plane, we can rotate the coordinate system until that condition is satisfied. Then we can choose the angle \cite{13}
\begin{equation}
\theta=\frac{\pi}{2}.
\end{equation}

We then have two remaining Killing vectors corresponding to the conserved quantity related to time translations and the magnitude of angular momentum. The conservation under time translations is obtained from the time-like Killing vector
\begin{equation}
K^\mu=(\partial_t)^\mu=(1,0,0,0).
\end{equation}

The Killing vector related to the angular momentum conservation is given by
\begin{equation}
R^\mu=(\partial_\phi)^\mu=(0,0,0,1).
\end{equation}

If we lower the index, then we obtain
\begin{equation}
K_\mu=(-e^{\nu(r)},0,0,0).
\end{equation}

Thus
\begin{equation}
K_\mu=\left(-\left(1-\frac{r_s}{r}-\frac{r^2}{3r_\varLambda^2}\right),0,0,0\right),
\end{equation}
\begin{equation}
R_\mu=(0,0,0,r^2).
\end{equation}

From Equation~(\ref{eq:CQa}), the two conserved quantities are
\begin{equation}    \label{eq:energy}
E=-K_\mu\frac{dx^\mu}{d\lambda}=e^{\nu(r)}\frac{dt}{d\tau},
\end{equation}
\begin{equation}   \label{eq:angmoment}
L=R_\mu\frac{dx^\mu}{d\lambda}=r^2\frac{d\phi}{d\tau},
\end{equation}
where $L$ is the magnitude of the angular momentum, and $\tau=\lambda$ is the proper time. Developing~explicitly Equation~(\ref{eq:KE}) for massive test particles, we obtain
\begin{equation}
-1=g_{00}\left(\frac{dt}{d\tau}\right)^2+g_{ii}\left(\frac{dx^i}{d\tau}\right)^2.
\end{equation}

If we introduce the S-dS metric given by Equation~(\ref{eq:Sdsm}) and additionally we use the results~(\ref{eq:energy}) and~(\ref{eq:angmoment}), then we get
\begin{equation}   \label{eq:Moveq}
\frac{1}{2}\left(\frac{dr}{d\tau}\right)^2+\frac{L^2}{2r^2}-\frac{r_sL^2}{2r^3}-\frac{r_s}{2r}-\frac{1}{6}\frac{r^2}{r_\varLambda^2}=\frac{1}{2}\left(E^2+\frac{L^2}{3r_\varLambda^2}-1\right)=C,
\end{equation}
where $C$ is a constant depending on the initial conditions of motion, we can define the effective potential as
\begin{equation}   \label{eq:effpotaa}
U_{eff}(r)=-\frac{r_s}{2r}-\frac{1}{6}\frac{r^2}{r_\varLambda^2}+\frac{L^2}{2r^2}-\frac{r_sL^2}{2r^3}.
\end{equation}

Then Equation~(\ref{eq:Moveq}) is equivalent to
\begin{equation}   \label{eq:Moveq2}
\frac{1}{2}\left(\frac{dr}{d\tau}\right)^2+U_{eff}(r)=\frac{1}{2}\left(E^2+\frac{L^2}{3r_\varLambda^2}-1\right)=C.
\end{equation}

The first term on the right-hand side of Equation~(\ref{eq:effpotaa}) is the Newtonian gravitational potential, the second term is the $\Lambda$ contribution (it reproduces a repulsive effect), the third term is the centrifugal force and it takes the same form in Newtonian gravity and General Relativity. The last term is the General Relativity correction to the effective potential. This last contribution is important at short scales, i.e.,~comparable to the gravitational radius $2GM$. Normally $r_\Lambda>>r_s$ and if $r_\Lambda\approx r_s$, then the black-hole is near its maximum mass value before it becomes a naked singularity. The event horizons for the black-hole solution are obtained by the condition
\begin{equation}   \label{eq:ov}
0=1-\frac{r_s}{2r}-\frac{r^2}{3r_\varLambda^2},
\end{equation}
and they are given explicitly by
\begin{equation}  \label{eq:7} 
r_{CH}=-2r_\varLambda \cos\left(\frac{1}{3}\left(\cos^{-1}\left(\frac{3r_s}{2r_\varLambda}\right)+2\pi\right)\right),
\end{equation}
\begin{equation*}
r_{BH}=-2r_\varLambda \cos\left(\frac{1}{3}\left(\cos^{-1}\left(\frac{3r_s}{2r_\varLambda}\right)+4\pi\right)\right),
\end{equation*}
where $r_{CH}$ is the Cosmological Horizon and $r_{BH}$ is the Black Hole event horizon. The S-dS solution previously defined in Equation~(\ref{eq:Sdsm}) is only valid for the region between the two event horizons such that we can still define static observers. If $r_\Lambda\sim r_s$, then $r_{CH}\sim r_{BH}$ and then our coordinate system is inappropriate  since the region between the two event horizons is very small (negligible) \cite{3}. Then~it is impossible to define static observers in most of the spacetime. Under the condition $r_\Lambda>>r_s$, the~results obtained in (\ref{eq:7}) can be expanded as \cite{2}
\begin{equation}   \label{eq:7r}
r_{CH}\approx \sqrt{3}r_\Lambda-\frac{1}{2}r_s,\;\;\;\;\;r_{BH}\approx r_s+\frac{1}{6}\frac{r_s^4}{r_\Lambda^3}.
\end{equation}   

\section{Circular Orbit Conditions for the Effective Potential with $\varLambda$}   \label{eq:Critical}

We can now obtain the critical points for the effective potential obtained from Equation~(\ref{eq:effpotaa}). For~illustration purposes, we start the analysis with the case where the Cosmological Constant vanishes ($\varLambda=0$). This case is important because the condition $\Lambda\neq0$ will only affect the local physics at astrophysical scales or larger. Then any result at the solar system scale for example, will not be affected by the condition $\Lambda\neq0$. We know that the effective potential for a massive test particle is given by \cite{13}
\begin{equation}   \label{eq:EfPot}
U_{eff}(r)_{\varLambda=0}=-\frac{r_s}{2r}+\frac{L^2}{2r^2}-\frac{r_s L^2}{2r^3}.
\end{equation}

Taking the derivative with respect to r of this equation we then obtain
\begin{equation}   \label{eq:Derivative of potential}
\frac{dU_{eff}(r)_{\varLambda=0}}{dr}=\frac{r_s}{2r^2}-\frac{L^2}{r^3}+\frac{3r_sL^2}{2r^4}=0,
\end{equation}
which can be expressed as
\begin{equation}   \label{eq:Poly2}
r_c^2-\frac{2L^2}{r_s}r_c+3L^2=0,
\end{equation}
where we have defined $r_c$ as the distance at which the orbits for a test particle are circular. The solutions for the previous equation are
\begin{equation}   \label{eq:circ orbit}
r_c=\frac{L^2\pm \sqrt{L^4-3r_s^2L^2}}{r_s}.
\end{equation}

If $L \to \infty$, then we get
\begin{equation}   \label{eq:atoinf}
r_c\approx\frac{L^2\pm L^2\left(1-\frac{3r_s^2}{2L^2}\right)}{r_s}=\left(\frac{2L^2}{r_s},\frac{3}{2}r_s\right).
\end{equation}

The first result corresponds to the stable equilibrium \cite{13} and the second circular orbit scale is proportional to the Schwarzschild radius and it corresponds to an unstable equilibrium. The minimum angular momentum in order to get bound orbits is obtained as the discriminant of Equation~(\ref{eq:circ orbit}) is zero, in such a case
\begin{equation}   \label{eq:Min angular mom}
L_{min}=\sqrt{3}r_s.
\end{equation}

If the angular momentum takes this value, the two distances for circular orbits obtained in Equation~(\ref{eq:circ orbit}) become to be the same and they correspond to a saddle point condition defined by
\begin{equation}   \label{eq:saddp}
r_{cx}=3r_s.
\end{equation}

Since this result corresponds to the saddle point condition, it can be found equivalently if we solve simultaneously the equations obtained from the vanishing condition for the first and second derivatives of the effective potential defined by Equation~(\ref{eq:EfPot}). If we replace the second solution obtained in Equation~(\ref{eq:atoinf}) $\left(r_c=\frac{3}{2}r_s\right)$ inside Equation~(\ref{eq:Derivative of potential}), we obtain
\begin{equation}
\frac{dU_{eff}(r)}{dr}_{\varLambda=0}=\frac{2}{9r_s}-\frac{8L^2}{27r^3_s}+\frac{8L^2}{27r^3_s}=\frac{2}{9r_s}.
\end{equation}

Then the scale $r_c=\frac{3}{2}r_s$ is not an exact solution of Equation~(\ref{eq:Poly2}). This scale becomes to be an~exact solution if the approximation $L>>r_s$ is satisfied. \mbox{In such a case}, the term corresponding to the Newtonian contribution and given by $\frac{r_s}{r^2}$ is negligible in Equation~(\ref{eq:Derivative of potential}) at short scales $r$. At~the moment of evaluating the case with $\Lambda\neq0$, the short distance behavior of the potential will be the same as in the present case because the $\Lambda$ effects will be only relevant at scales larger than $r_0=(3GMr_\Lambda^2)^{1/3}$. In~massive gravity theories, the Vainshtein scale plays an analogous role as we will verify later. Then~under the previous condition, for short scales ($r\to r_s$), the effective potential can be approximated as follows
\begin{equation}   \label{eq:tepapp}
U_{eff}(r)_{\varLambda=0}\approx \frac{L^2}{2r^2}-\frac{r_s L^2}{2r^3}.
\end{equation}

As a proof of this statement, we can take the derivative with respect to $r$ from the previous~equation
\begin{equation}   \label{eq:doepfsvod}
\frac{dU_{eff}(r)}{dr}_{\varLambda=0}\approx -\frac{L^2}{r^3}+\frac{3r_sL^2}{2r^4}.
\end{equation}

\begin{figure}
	\centering
		\includegraphics[width=15cm, height=8cm]{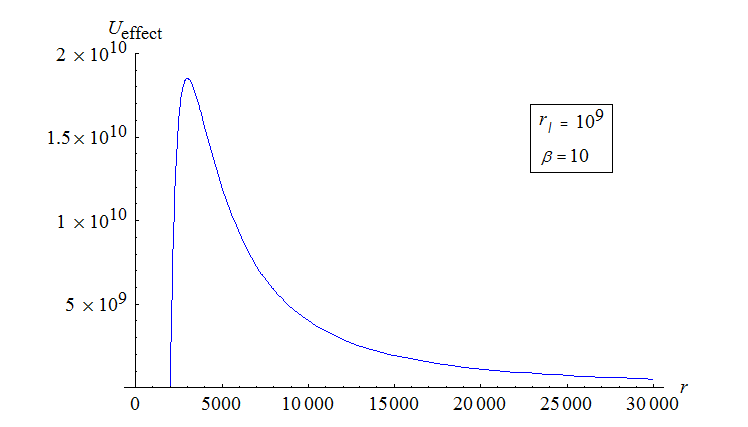}
	\caption{The effective potential affecting the motion of a test particle at scales where the scale $GM$ is important. Here $\beta$ is the ratio between the maximum angular momentum scale (to be defined later) and the ordinary angular momentum here defined as $L=r_l$.}  %Please revise the format of the numbers in the y-axis, such as ``5. ×'', which should be ``5 ×''. A. I have corrected this.
	\label{fig:eff}
\end{figure}

If we set this result to zero, then we obtain
\begin{equation}   \label{eq:mpwdz}
r_1=\frac{3}{2}r_s,
\end{equation}
which corresponds to the second solution given in Equation~(\ref{eq:atoinf}). The Figure~\ref{fig:eff} illustrates the effective potential curve for scales where the gravitational radius $GM$ is the dominant one. The second circular orbit located at $r_2=\frac{2L^2}{r_s}$ (see Equation~(\ref{eq:atoinf})) is not an exact solution of Equation~(\ref{eq:Derivative of potential}) neither. This can be seen by replacing $r_2$ in Equation~(\ref{eq:Derivative of potential}) and verifying that

\begin{figure}
	\centering
		\includegraphics[width=15cm, height=8cm]{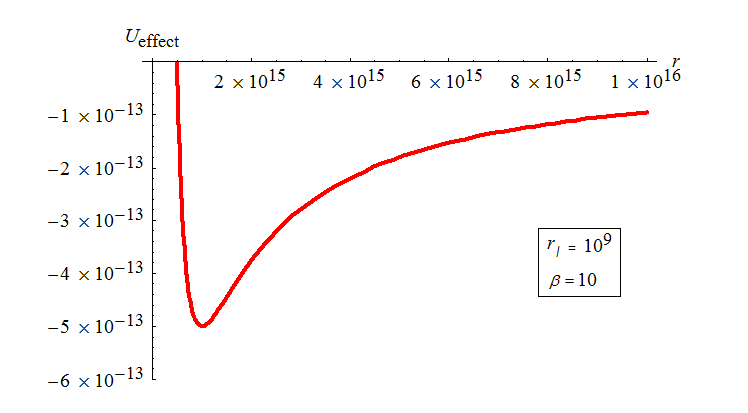}
	\caption{Effective potential valid for scales satisfying $r_s<<r<<r_0$ and $L>>r_s$ for a particular value of $\beta$. The shape of the figure will not change for different values of $\beta$. Here $r_l=L$ is the angular momentum scale.}
	\label{fig:Ueff1}
\end{figure}

\begin{equation}
\frac{dU_{eff}(r)_{\varLambda=0}}{dr}=\frac{3r^5_s}{32L^6},
\end{equation}
evaluated at $r=r_2$. In fact, $r_2$ becomes an exact solution only if the term $\frac{3r_sL^2}{2r^4}$ (GR contribution) from Equation~(\ref{eq:Derivative of potential}) is neglected. Then for large scales, under the assumption $L>>r_s$, as far as the cosmological constant effects are neglected, the effective potential ($\Lambda=0$) is defined by
\begin{equation}   \label{eq:epaldwol}
U_{eff}(r)_{\varLambda=0}\approx-\frac{r_s}{2r}+\frac{L^2}{2r^2},
\end{equation}
ignoring then the GR contribution which mix the scales $r_s$ with $L$. In order to verify this assumption, we can calculate the extremal condition for the potential (\ref{eq:epaldwol}), obtaining then
\begin{equation}
\frac{dU_{eff}(r)}{dr}_{\varLambda=0}\approx \frac{r_s}{2r^2}-\frac{L^2}{r^3}=0,
\end{equation}
from which we can obtain the result $r_2$ defined previously. The Figure \ref{fig:Ueff1} illustrates the effective potential curve at large scales under the condition $L>>r_s$. This portion of the curve corresponds to the stable orbit condition. This part of the effective potential will not suffer any modification when the cosmological constant effects are included.

\subsection{{\bf The Case with $\Lambda\neq0$}}

We can repeat the previous arguments. Then the full effective potential with the inclusion of the $\Lambda$ term is given by
\begin{equation}    \label{eq:Total eff}
U_{eff}(r)=-\frac{r_s}{2r}-\frac{r^2}{6r_\varLambda^2}+\frac{L^2}{2r^2}-\frac{r_s L^2}{2r^3}.
\end{equation}

At large scales, with $L>>r_s$, the previous potential can be approximated to
\begin{equation}    \label{eq:iccterm}
U_{eff}(r)\approx-\frac{r_s}{2r}-\frac{r^2}{6r_\varLambda^2}+\frac{L^2}{2r^2},
\end{equation}
where again we have ignored the GR correction term $-\frac{r_sL^2}{2r^3}$ due to the same arguments exposed previously. The condition $dU_{eff}(r)/dr=0$ in Equation~(\ref{eq:iccterm}) gives the condition 
\begin{equation}   \label{eq:forder}
r^4-\frac{3}{2}r_\varLambda^2 r_s r+3r_\varLambda^2L^2=0.
\end{equation}

This is a reduced fourth order polynomial as has been defined in Appendix~\ref{eq:fos}. We can compare with the standard reduced form defined by Equation~(\ref{eq:reduced}). After following a standard procedure, we~can define the discriminant for the quartic polynomial (\ref{eq:forder}) as follows 
\begin{equation}   \label{eq:Dfac}
D=-64r_\varLambda^6L^6+\frac{81}{64}r_\varLambda^8r_s^4,
\end{equation}
in agreement with the result defined in Equation~(\ref{eq:D}) in the corresponding appendix. This~discriminant plays an analogous role to the root square term in a quadratic polynomial equation as it is the case in Equation~(\ref{eq:circ orbit}). The type of solution obtained depends on the sign of the discriminant. In fact, it can be demonstrated that if $D<0$, then there are no real solutions. If $D>0$, we can get real solutions. Then $D=0$ represents a limit where we get the largest possible angular momentum such that still bound orbits are possible. The maximum angular momentum is defined by
\begin{equation}  \label{eq:Max angular mom}
L_{max}=\frac{{3}^{2/3}}{4}(r_s^2r_{\varLambda})^{1/3}.
\end{equation}

This condition is coincident with the result obtained in \cite{5}. The solutions for Equation~(\ref{eq:forder}), when~$D>0$, are defined by 
\begin{equation}   \label{eq:frwbimbto}
r_1^*=\frac{2L^2}{r_s}, \;\;\;\;\;\;\;\;\; r_2^*=4^{1/3}\sqrt{r_\varLambda L_{max}}-\frac{1}{2\beta^2}\sqrt{r_\varLambda L_{max}}.
\end{equation}

\begin{figure}
	\centering
		\includegraphics[width=15cm, height=8cm]{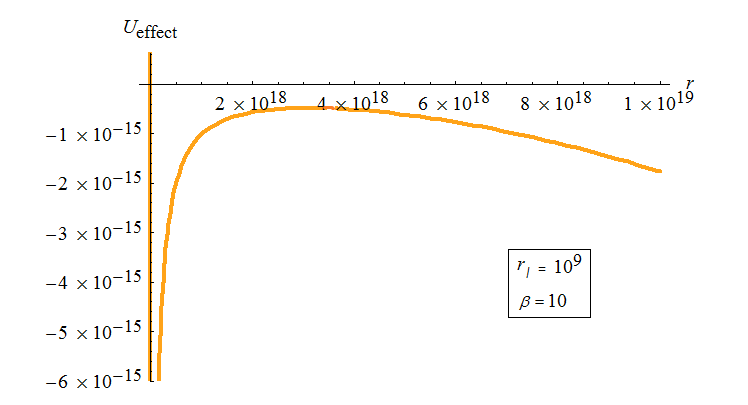}
	\caption{Effective potential scales of the order $r\approx r_0$ and $L>>r_s$ for some specific value of $\beta$.}
	\label{fig:Ueff2}
\end{figure}

The first solution is coincident with the second solution found for Equation~(\ref{eq:Poly2}). The second solution is new since it comes from the $\Lambda$ contribution, which was excluded previously. By using here the result (\ref{eq:Max angular mom}), we can express $r_2^*$ as
\begin{equation}   \label{eq:srwbimbto}
r_2^*=\left(\frac{3}{2}r_sr_\varLambda^2\right)^{1/3}-\frac{1}{4\beta^2}(3r_sr_\varLambda^2)^{1/3}=r_0',
\end{equation}
where $\beta\equiv L_{max}/L$. This solution corresponds to the local maximum illustrated in the Figure \ref{fig:Ueff2}. The condition $L=L_{max}$ is equivalent to $\beta=1$, which corresponds to a second saddle point condition different to the one defined previously for the result (\ref{eq:Min angular mom}). This new saddle point is located at large scales defined by 
\begin{equation}   \label{eq:spbomam}
r_1^*=r_2^*=\sqrt{r_\varLambda L_{max}}=r_x.
\end{equation}

If we use Equation~(\ref{eq:Max angular mom}) in the previous solutions, then we get
\begin{equation}    \label{eq:uniqueroot}
r_1^*=r_2^*=r_x=\frac{(3)^{1/3}}{2}(r_sr_\varLambda^2)^{1/3},
\end{equation}
which is the same result obtained in \cite{5} by using different methods.
%\vspace{-24pt}
  %This figure has not been referred to within the text of the manuscript. Please cite it. Please check the format of the numbers in the y-axis, such as ``5. ×'', which should be ``5 ×''.

\vspace{-12pt}
  %This figure has not been referred to within the text of the manuscript. Please cite it. Please check the format of the numbers in the y-axis, such as ``5. ×'', which should be ``5 ×''.

\section{The Role of $r_0(\beta=0)$ in the Black-Hole Thermodynamics in S-dS Space}   \label{eq:Momoko Cheung}

The black-hole surface gravity, can be defined  as \cite{3}
\begin{equation}   \label{eq:12}
\kappa_{BH, CH}=\left(\frac{(K^\mu\nabla_\mu K_\gamma)(K^\alpha\nabla_\alpha K^\gamma)}{-K^2}\right)^{1/2}_{r=r_{BH},r_{CH}}.
\end{equation}

The labels BH and CH, correspond to the Black Hole Horizon and the Cosmological Horizon respectively. From Equation (\ref{eq:7}), the two horizons are the same when the mass of the Black Hole takes its maximum value given by

\begin{equation}   \label{eq:8}
M_{max}=\frac{1}{3}\frac{m_{pl}^2}{m_\varLambda},
\end{equation}
where $m_{pl}$ corresponds to the Planck mass and $m_\Lambda=\sqrt{\Lambda}$. If the mass of a Black Hole is larger than the value given by Equation~(\ref{eq:8}), then we have a naked singularity. As~$M=M_{max}$, the two event horizons are equally defined as $r_{BH}=r_{CH}=r_\Lambda=\frac{1}{\sqrt{\Lambda}}$. Then a thermodynamic equilibrium is established. As~has been explained in \cite{3}, if~$M\to M_{max}$, then the metric defined in Equation~(\ref{eq:Sdsm}) obeys the condition $e^{\nu(r)}\to0$ everywhere. Then~this coordinate system becomes inappropriate. In agreement with the results obtained in \cite{16}, we can select a~better coordinate system by defining
\begin{equation}   \label{eq:9}
9M^2\Lambda=1-3\omega^2,   \;\;\;\;\;0\leq\omega\ll1,
\end{equation} 
where $\omega$ is a parameter related to the mass of the black-hole. In these coordinates, the degenerate case (when the two horizons become the same), corresponds to $\omega\to0$. We must then define the new radial and the new time coordinates to be
\begin{equation}   \label{eq:10}
\tau=\frac{1}{\omega\sqrt{\Lambda}}\psi,\;\;\;\;\; r=\frac{1}{\sqrt{\Lambda}}\left(1-\omega cos\chi-\frac{1}{6}\omega^2\right).
\end{equation}

In these coordinates, the Black Hole horizon corresponds to $\chi=0$ and the Cosmological horizon to $\chi=\pi$ \cite{3}. Then the metric (\ref{eq:Sdsm}) can be expressed as
\begin{equation}   \label{eq:11}
ds^2=-r_\Lambda^2\left(1+\frac{2}{3}\omega cos \chi\right)sin^2\chi d\psi^2+r_\Lambda^2\left(1-\frac{2}{3}\omega cos\chi\right)d\chi^2+r_\Lambda^2(1-2\omega cos\chi)d\Omega_2^2,
\end{equation}  
after a standard coordinate transformation. The previous metric has been expanded up to first order in $\omega$. Equation~(\ref{eq:11}) is then the appropriate metric to be used as the mass of the Black Hole is near to its maximum value defined by Equation~(\ref{eq:8}). For any value taken by the mass of the black-hole, the~normalization of the time-like Killing vector corresponding to the conservation under time-translations is defined by
\begin{equation}   \label{eq:16}
\gamma_t=\left(1-\left(\frac{3r_s}{2r_\Lambda}\right)^{2/3}\right)^{-1/2},
\end{equation}
with $\gamma_t$ being the normalization factor for the time-like Killing vector defined by
\begin{equation}   \label{eq:13}
K=\gamma_t\frac{\partial}{\partial t}.
\end{equation}

In an asymptotically flat space, it is standard to define $\gamma_t\to1$ when $r\to\infty$. However, for the S-dS case, since we can only define the static observer at the scale $r_0$, then it is standard to normalize the time-like Killing vector as in Equation~(\ref{eq:13}) with the $\gamma_t$ factor defined by Equation~(\ref{eq:16}). For the case with $\beta\neq0$, where the angular momentum effects for the observer will appear, the corresponding corrections have to be done for the previous normalization factor. When the mass of the black hole satisfies $M=M_{max}$, then $\omega\to0$ in Equation~(\ref{eq:9}) and the black-hole temperature takes its minimum possible value defined by
\begin{equation}   \label{eq:19}
2\pi T_{min=}\kappa_{min}^{BH}=\frac{1}{r_\Lambda},
\end{equation}   
where $\kappa$ is the surface gravity. Later we will see how in massive gravity theories, an~analogous expression for the black-hole temperature appears. For any value taken by the black-hole mass, the~black-hole temperature is defined by  
\begin{equation}
\kappa_{BH, CH}=\frac{1}{2\sqrt{U(r_0)}}\left\vert\frac{\partial U}{\partial r}\right\vert_{r=r_{BH}, r_{CH}},
\end{equation}  
where $r_0$ is the location of the static observer inside the S-dS solution. Other relevant analysis for the black-hole temperature in the S-dS solution in GR have been done in \cite{3}.

\section{The Schwarzschild De-Sitter Solution in Massive Gravity} \label{eq:Final1}

Here we will make a review of the derivation of the S-dS solution inside massive gravity. We then start by obtaining the field equations from the variation of the action
\begin{equation}   \label{eq:b1}
S=\frac{1}{2\kappa^2}\int d^4x\sqrt{-g}(R+m^2U(g,\phi)).
\end{equation}
with the effective potential depending on two free parameters as
\begin{equation}   \label{eq:b2}
U(g,\phi)=U_2+\alpha_3 U_3+\alpha_4U_4.
\end{equation}

By convenience, here we redefine the free-parameters as follows
\begin{equation}
\alpha=1+3\alpha_3,\;\;\; \beta=3(\alpha_3+4\alpha_4).
\end{equation}

Here in addition
\begin{equation}   \label{eq:b3}
U_2=Q^2-Q_2,\;\;\;\;\;U_3=Q^3-3QQ_2+2Q_3,\;\;\;\;\;U_4=Q^4-6Q^2Q_2+8QQ_3+3Q_2^2-6Q_4,
\end{equation}
and
\begin{equation}   \label{eq:b6}
Q=Q_1\;\;\;\;\;Q_n=Tr(Q^n)^\mu_{\;\;\nu},
\end{equation}
\begin{equation}   \label{eq:b7}
Q^\mu_{\;\;\nu}=\delta^\mu_{\;\;\nu}-M^\mu_{\;\;\nu},
\end{equation}
\begin{equation}   \label{eq:b8}
(M^2)^\mu_{\;\;\nu}=g^{\mu\alpha}f_{\alpha\nu},
\end{equation}
\begin{equation}   \label{eq:b9}
f_{\mu \nu}=\eta_{ab}\partial_\mu\phi^a\partial_\nu\phi^b,
\end{equation}
and then the field equations obtained from the variation of Equation~(\ref{eq:b1}) are defined as
\begin{equation}   \label{eq:b10}
G_{\mu \nu}=-m^2X_{\mu\nu}.
\end{equation}

The energy momentum tensor defined by
\begin{equation}   \label{eq:b11}
X_{\mu \nu}=\frac{\delta U}{\delta g^{\mu \nu}}-\frac{1}{2}Ug_{\mu \nu}.
\end{equation}

We can constraint the background solution to behave as the standard S-dS solution of GR. In such a case, the following result has to be satisfied
\begin{equation}   \label{eq:b21}
m^2X_{\mu\nu}=\Lambda g_{\mu \nu}.
\end{equation}

Then we get the solution
\begin{equation}   \label{eq:drgt metric2}
ds^2=g_{tt}dt^2+g_{rr}dr^2+g_{rt}(drdt+dtdr)+{S_0^2}r^2d\Omega_2^2,
\end{equation}
with
\begin{eqnarray}   \label{eq:drgt metric}
g_{tt}=-f(S_0r)(\partial_tT_0(r,t))^2,\;\;\;\;\;g_{rr}=-f(S_0r)(\partial_rT_0(r,t))^2+\frac{S_0^2}{f(S_0r)},\nonumber\\
g_{tr}=-f(S_0r)\partial_tT_0(r,t)\partial_rT_0(r,t),
\end{eqnarray}
where $f(S_0 r)=1-\frac{2GM}{S_0 r}-\frac{1}{3}\Lambda S_0^2 r^2$. In this previous solution, all the degrees of freedom are inside the dynamical metric. The fiducial metric in this case is simply defined by Minkowski. It was found in~\cite{Kodama, Kodama-2} that the generic solution defined by Equation~(\ref{eq:drgt metric}), can be classified in two families depending on the relation between the two free-parameters of the theory. The solutions are then defined as

\subsection{Solution with a Degenerate Vacuum} 

In this particular solution, the condition $\beta=\alpha^2$ is satisfied. Then the St\"uckelberg function defined by $T_0(r,t)$ is arbitrary. For this case, the scale factor is defined as
\begin{equation}
S_0=\frac{\alpha}{1+\alpha},
\end{equation}  
and the cosmological constant, due to the constraint imposed in Equation~(\ref{eq:b21}) has to satisfy the~condition
\begin{equation}
\Lambda=\frac{m^2}{\alpha},
\end{equation}
as can be easily verified. 

\subsection{Solution with Single Vacuum}

For this solution, it was found that the function $T_0(r,t)$ is constrained to behave as a solution of the following equation
\begin{equation}   \label{stcons}
(T_0')^2=\frac{1-f(S_0r)}{f(S_0 r)}\left(\frac{S_0^2}{f(S_0 r)}-\dot{T}_0^2\right).
\end{equation} 

Note that the solution for this equation has the Finkelstein form as has been demonstrated in \cite{Kodama, Kodama-2}. In general the function $T_0(r,t)$ has the following solution in this case
\begin{equation}
T_0(r,t)=S_0t+A(r,t),
\end{equation} 
where $A(r,t)$ is some function depending on spacetime. For the stationary case, $A(r,t)$ only has dependence on $r$. In order to satisfy the constraint (\ref{stcons}), here we have \cite{Kodama}
\begin{equation}
A(r,t)=A(r)=\pm\int^{S_0r}\left(\frac{1}{f(u)}-1\right).
\end{equation} 

For this solution in addition, the scale factor is defined as
\begin{equation}
S_0=\frac{\alpha+\beta\pm\sqrt{\alpha^2-\beta}}{1+2\alpha+\beta},
\end{equation} 
and the cosmological constant for this case is defined as
\begin{equation}
\Lambda=-m^2\left(1-\frac{1}{S_0}\right)\left(2+\alpha-\frac{\alpha}{S_0}\right).
\end{equation}

Note that independent of the type of solution, the final result can be expressed as in Equation~(\ref{eq:drgt metric}). In the analysis of the motion of a test particle, it is enough to consider the situation where the solution is expressed in the form (\ref{eq:drgt metric}). 

\subsection{The Effective Potential in dRGT Massive Gravity}     

Having the generic solution for the spherically symmetric case, then we can proceed to calculate the equations of motion for a test particle moving under the influence of a spherically symmetric source inside the massive gravity formulation. Our derivation will be based in the solution formulated in Equation~(\ref{eq:drgt metric}) for a general St\"uckelberg function $T_0(r,t)$. The equation to be analyzed is
\begin{equation}
-1=g_{00}\left(\frac{dt}{d\tau}\right)^2+g_{ii}\left(\frac{dx^i}{d\tau}\right)^2+2g_{0i}\left(\frac{dt}{d\tau}\right)\left(\frac{dx^i}{d\tau}\right).
\end{equation}

Note that in this case the non-diagonal terms in the $t-r$ components will be relevant since we cannot gauge them away as in the GR case. Here instead of defining a conserved quantity in the direction of $t$, we define a  conserved quantity in the direction of the St\"uckelberg function $T_0(r,t)$ \mbox{as follows}
\begin{equation}   \label{conservation}
\varepsilon=-K_{T_0(r,t)}\frac{dT_0(r,t)}{d\lambda}=-K_{T_0(r,t)}\left(\dot{T}_0(r,t)\frac{dt}{d\lambda}+T_0'(r,t)\frac{dr}{d\lambda}\right),
\end{equation}
{with $K_{T_0(r,t)}=\gamma_tf(S_0r)\left(1,0,0,0\right)$. Here $\gamma_t$ is a normalization factor, analogous to the one introduced in \cite{3}. This previous relation is complemented with} the standard conserved angular momentum already defined in Equation~(\ref{eq:angmoment}). Then in this case we obtain the following equation of motion for \mbox{a test particle}
\begin{equation}   \label{eq:potra}
\frac{1}{2}\left(\frac{d(S_0 r)}{dr}\right)^2-\frac{GM}{S_0r}-\frac{1}{6}\Lambda (S_0 r)^2+\frac{L^2}{2(S_0r)^2}-\frac{GM L^2}{(S_0r)^3}=A=\frac{1}{2}\left(\varepsilon^2-1+\frac{1}{3}\Lambda L^2\right).
\end{equation}

Here $A$ is a constant of motion. Then the potential will behave exactly as in the GR case, although~there is a~big difference in the behavior of the kinetic terms. In fact, for the case of massive gravity, the quantity associated to the time-translations is not conserved as in the case of GR. This can be observed from Equation~(\ref{conservation}). Here the conservation is only well defined in the direction of $T_0(r,t)$. Then what is in reality conserved is a combination of a quantity defined in the direction of the ordinary time $t$ and a quantity defined in the spatial direction ($r$). Then although the potential term will have exactly the same behavior as in the case of GR, the combination of the kinetic term representing radial translations, together with the kinetic term representing temporal translations, will be combined such that their superposition (up to some factors), is a constant of motion. Then there is no conservation under time-translations in the usual sense in this case. This is a big difference with respect to the GR case, where there is conservation under ordinary time-translations. Another difference of the present case with respect to GR is how we define the scale $r_0$ for the potential (\ref{eq:potra}). In this case, such scale will depend on the graviton mass and on the two free-parameters of the theory through the constants $\Lambda$ and $S_0$ \cite{Kodama}. It is not difficult to visualize that $r_0=r_V$ corresponds to the Vainshtein scale and it can be calculated in the same way as we did for the case of $r_0$ in GR. Then we can still use one of the solutions of the quartic equation defined in Equation~(\ref{eq:forder}) as the Vainshtein scale. The Vainshtein scale then~becomes
\begin{equation}
r_V\approx\frac{1}{S_0}\left(\frac{3GM}{\Lambda}\right)^{1/3},
\end{equation}     
up to some small corrections due to the angular momentum, analogous to the corrections found in Equation~(\ref{eq:srwbimbto}) for the case of ordinary gravity. Note that in this case $S_0$ and $\Lambda$ will take different values depending on the relations between the two free-parameters.

\subsubsection{The Case of One Free-Parameter}

In this case, $S_0=\frac{\alpha}{1+\alpha}$ and $\Lambda=\frac{m^2}{\alpha}$. Then
\begin{equation}
r_V=\left(\frac{\alpha}{1+\alpha}\right)\left(\frac{3GM\alpha}{m^2}\right)^{1/3}. 
\end{equation}

This scale clearly depends on the graviton mass \cite{Kodama}.

\subsubsection{The Case of Two Free-Parameters}

In this case $S_0$ is a function of the two free-parameters as it has been defined in \cite{Kodama}. In addition
\begin{equation}
\Lambda=-m^2\left(1-\frac{1}{S_0}\right)\left(2+\alpha-\frac{\alpha}{S_0}\right).
\end{equation}  

Then $r_V$ can be defined correspondingly. 

\section{Black-Hole Thermodynamics Inside the Non-Linear Formulation of Massive Gravity}   \label{BHTINLMG}

The expression for the black-hole temperature inside the non-linear formulation of massive gravity for the S-dS solution, will not be different to the one defined previously for the GR case as far as we use the Killing vector in the direction of $T_0(r,t)$ instead of using the ordinary time-like Killing vector as follows 
\begin{equation}   \label{RealIvan2}
K^{T_0(r,t)}=\gamma_t\frac{\partial}{\partial T_0(r,t)}=\gamma_t\left(\frac{1}{\dot{T}_0(r,t)}K^t+\frac{1}{T'_0(r,t)}K^r\right).
\end{equation}

{Here we constraint $K^{T_0(r,t)}=(1,0,0,0)$. In addition, $K^r$ is the unit vector which marks the radial component of $K^{T_0(r,t)}$. $\gamma_t$ is the normalization factor}. We can now normalize this Killing vector as we did with the time-like Killing vector $K^t$ for the case of GR. Here however, the static observer is located at the scale $r_V$ corresponding to the Vainshtein scale and in order to agree with the GR results, this observer must define his/her time-coordinate in the direction of $T_0(r,t)$. This imposes some extra-constraints in the type of observers able to define the black-hole temperature as in GR. Any other observer defining a different notion of time, will disagree with the results obtained in GR, even if this observers is also located at the same scale. The difference will be more evident for observers located at larger scales than $r_v$ (in massive gravity) or $r_0$ (in GR). The notion of time then is an important concept in massive gravity and it is related to how we define the vacuum and the conservation laws. Here we normalize the Killing vector in the direction of $T_0(r,t)$ as     
\begin{equation}   \label{UNI2}
K^2=K_\mu K^\mu=g_{\mu \nu}K^\mu K^\nu=\gamma_t^2(g_{T_0T_0})\vert _{r=r_V}=-\gamma_t^2\left(1-\left(\frac{3r_s}{2r_\Lambda}\right)^{2/3}\right)=-1.
\end{equation}

{Note that here since we are considering the Killing vector in the direction $T_0(r,t)$, then we have to consider the metric component $g_{T_0 T_0}$ in Equation~(\ref{UNI2}). This component can be obtained from the result (\ref{eq:drgt metric}) if we rewrite the metric (\ref{eq:drgt metric2}) as}
\begin{equation}   \label{UNI3}
ds^2=g_{T_0T_0}dT_0^2(r,t)+g_{rr}dr^2+S_0^2r^2d\Omega_2^2,
\end{equation}
{where $dT_0(r,t)=\dot{T}_0(r,t)dt+T_0'(r,t)dr$. If we compare the results (\ref{eq:drgt metric2}) and (\ref{eq:drgt metric}) with Equation~(\ref{UNI3}), then we conclude that $g_{T_0 T_0}=-f(S_0 r)=-\left(1-\frac{2GM}{S_0 r}-\frac{1}{3}\Lambda S_0^2r^2\right)$. This justifies the previous results. Then we conclude that the normalization factor is} 
\begin{equation}
\gamma_t=\left(1-\left(\frac{3r_s}{2r_\Lambda}\right)^{2/3}\right)^{-1/2},
\end{equation}
{consistent with the result obtained inside GR in Equation~(\ref{eq:16}). Note however that here we are normalizing the Killing vector in the direction $T_0(r,t)$ instead of the Killing vector in the time-direction $t$. In addition the normalization is done with respect to the observers located at $r=r_V$, namely, the~Vainshtein scale. If we want to get the constraints for the St\"uckelberg function $T_0(r,t)$, consistent with the normalization (\ref{UNI2}) and with the conservation law (\ref{conservation}), we have to expand Equation~(\ref{UNI2}) in the alternative form}
\begin{equation}   \label{UNI4}
K^2=K_\mu K^\mu=g_{\mu \nu}K^\mu K^\nu=\gamma_t^2\left(g_{tt}(K^t)^2+g_{rr}^{T_0(r,t)}(K^r)^2+2g_{tr}K^tK^r\right)=-1.
\end{equation} 

Here $g_{rr}^{T_0(r,t)}$ corresponds to the radial component of the metric living in the subspace formed by the vector $K^{T_0(r,t)}$. In other words, the components of the Killing vector in the direction of the St\"uckelberg function $T_0(r,t)$ can be defined as
\begin{equation}
K^{T_0(r,t)}=\left((a,b), 0, 0, 0\right)=(1,0,0,0). 
\end{equation}

Here $a$ and $b$ define the components of $K^{T_0(r,t)}$ inside a 2-dimensional subspace formed by $K^t$ and $K^r$ taken as unit vectors. This is the subspace where the conservation law (\ref{conservation}) is valid. Then in general we can express $K^{T_0(r,t)}=aK^t+bK^r$, consistent with the definition (\ref{RealIvan2}). Then here $g_{rr}^{T_0(r,t)}=-f(S_0r)T_0'^2(r,t)$ in Equation~(\ref{UNI4}). Taking into account this analysis, together with the solutions (\ref{eq:drgt metric}), then we find that Equation~(\ref{UNI4}) can be simplified to
\begin{equation}   \label{UNI5}
K^2=-f(S_0r_V)\gamma_t^2\left(\dot{T}_0^2(r,t)+T_0'(r,t)^2+2\dot{T}_0T_0'(r,t)\right)=-1,
\end{equation}  
taking into account that inside the 2-dimensional subspace expanded by $K^{T_0(r,t)}$, $K^t=(1,0)$ and $K^r=(0,1)$. If this previous result is consistent with Equation~(\ref{UNI2}), then the following constraint has to be satisfied
\begin{equation}   \label{RealIvan3}
T_0'^2(r,t)+\dot{T}_0^2(r,t)+2T_0'(r,t)\dot{T}_0(r,t)=1.
\end{equation}

This expression can be reduced as
\begin{equation}
\left(T_0'(r,t)+\dot{T}_0(r,t)\right)^2=1.
\end{equation}

This result is consistent with the conservation law defined in Equation~(\ref{conservation}) with the appropriate definitions for $K^t$ and $K^r$. Within this scenario, the definition of surface gravity for black-holes in Massive gravity agrees with the standard definition of GR. This is true as far as the observers are constrained to move in the spacetime defined by the condition (\ref{RealIvan3}). This condition is general and it applies for both cases, namely, for the case where the vacuum is degenerate or for the case where the vacuum is single, taking into account that massive gravity is in essence a $\sigma$-model. The St\"uckelberg trick as has been used in this paper can be found in \cite{Gaba3}.  % References [31] is cited after [28] directly, which is out of order. Please check whether [29,30] is missing and please renumber the references so they appear in sequential numerical order.
Note that the observers can define any location in space $r$ or any notion of time $t$. In other words, they can move arbitrarily. However, those observers satisfying the constraint (\ref{RealIvan3}), will agree with the results of GR. For general motion, the results obtained from GR will disagree with the results obtained in massive gravity.        

\section{The Number of Propagating Degrees of Freedom and Lorentz Violation}   \label{Finalsection}

The previous analysis has been based in solutions with non-trivial St\"uckelberg functions. In some scenarios this can be understood as Lorentz violating theories \cite{Rubakov, Rubakov-2}.   % References [34] is cited after [31] directly, please renumber the references so they appear in sequential numerical order.
The Lorentz violation is, however, at the background level due to the non-triviality of the St\"uckelberg functions. This can be considered as spontaneous symmetry breaking, since the action itself still respects the violated symmetries. In the previous analysis, the number of degrees of freedom can fluctuates from two to five depending on how the observers define the notion of time. If~an~observer defines the time arbitrarily, then he/she will describe five degrees of freedom in general. On the other hand, the observers defining the time in agreement with the St\"uckelberg function $T_0(r,t)$, will describe two degrees of freedom. However,~if~we describe a black-hole with spherical symmetry, then we will not have tensor perturbations and we can just limit the analysis to three degrees of freedom in four dimensions. This can be perceived better if we expand the Lagrangian density in the action (\ref{eq:b1}) up to second order. For simplicity we will work in a~free falling frame where we can ignore the scales of gravity. This will help us to make a~cleaner analysis. The free-falling condition here is different with respect to the one in GR because the dynamical metric in a free-falling frame does not necessarily goes to Minkowski, unless the St\"uckelberg function is trivial. By keeping the stationary condition, in a free-falling frame and for an arbitrary St\"uckelberg function, we get the dynamical
\begin{equation}   \label{eq:acalanojo234}
ds^2=S_0^2\left(-dt^2+dr^2\left[1-\left(\frac{T_0'(r,t)}{S_0}\right)^2\right]-2\frac{T_0'(r,t)}{S_0}dtdr+r^2d\Omega^2\right).
\end{equation}

Here we will consider the non-trivial contribution $T_0'(r,t)$ as a radial fluctuation of the St\"uckelberg function. In this special case $\dot{T}_0(r,t)=S_0$ and $T_0'(r,t)$ is arbitrary. In general we can assume $T_0(r,t)=S_0t+A^t(r,t)$. If we consider $A^t(r,t)/S_0t$ small, then the radial fluctuations of the St\"uckelberg function inside the 2-dimensional subspace correspond to perturbations. Such fluctuations live inside the space where we defined the Killing vector $K^{T_0(r,t)}$. In general, the metric (\ref{eq:acalanojo234}) is a consequence of the St\"uckelberg trick applied as 
\begin{equation}   \label{eq:stuefashio}
g_{\mu\nu}=\left(\frac{\partial Y^\alpha}{\partial x^\mu}\right)\left(\frac{\partial Y^\beta}{\partial x^\nu}\right)g'_{\alpha\beta},
\end{equation}  
with $Y^0(r,t)=T_0(r,t)$ and $Y^r=S_0r$. If we expand $Y^\alpha(x)=Sx^\alpha+A^\alpha(x)$, then $A^t(x)$ is non-trivial and $A^r(x)=0$ in the example developed here. By considering $A^\alpha(x)$ as perturbations, the dynamical metric would be Minkowski at the background level. The kinetic terms of the action will not change under the presence of the extra-degrees of freedom because they enter via the St\"uckelberg trick defined in Equation~(\ref{eq:stuefashio}). This is the case because the St\"uckelberg trick looks like (but is completely different) a coordinate transformation from the GR perspective. Then we only have to worry about the perturbations of the massive action. In a free-falling frame, up to second order, the massive action is defined~as
\begin{eqnarray}   \label{eq:acalanojo2miau}
\sqrt{-g}U(g,\phi)\approx \left(1+\frac{1}{2}h-\frac{1}{4}h^\alpha_{\;\;\beta}h^\beta_{\;\;\alpha}+\frac{1}{8}h^2\right)\left(\frac{2+6\alpha(1+\alpha)}{(1+\alpha)^4}\right)\nonumber\\
-\left(1+\frac{1}{2}h\right)\left(-\frac{h}{1+\alpha}+\frac{2T_0'(r,t)(1+\alpha)^2}{\alpha^3}h_{0r}-\frac{T_0'(r,t)^2(1+\alpha)^3}{\alpha^4}\right)+...     
\end{eqnarray}

We can see that the radial fluctuations of the time-component of the St\"uckelberg function affects the number of degrees of freedom perceived by an observer. If $T_0(r,t)$ becomes trivially the ordinary time coordinate, then the number of degrees of freedom is reduced to the ordinary case of GR taking into account that the massive term will become an ordinary cosmological constant. In conclusion, the number of degrees of freedom described by an observer will depend on the way how he/she defines the time coordinate with respect to the St\"uckelberg function. An observer defining $T_0(r,t)\backsim t$, will~perceive $T_0'(r,t)=0$ and will recover GR. Any other observer will perceive more degrees of freedom. Note that here for simplicity $\dot{T}_0(r,t)=S_0$. A non-trivial contributions coming from $\dot{T}_0(r,t)$ will also affect the number of degrees of freedom perceived by observers. In addition, if we evaluate the second derivatives with respect to the graviton field in Equation~(\ref{eq:acalanojo2miau}), we will obtain a mass matrix. In such a case, the eigenvalues of the matrix would correspond to the masses of each mode, namely, $m_{00}$, $m_{ij}$, etc. These are the mass fluctuations of the modes and they will depend on the non-trivial contributions of the St\"uckelberg function. In any other frame different to the free-falling one, the scales of gravity will appear but the analysis will be qualitatively the same. A similar analysis has been developed in \cite{Sigma, Sigma-2} where massive gravity was analyzed as a gravitational $\sigma$-model.   

\section{Conclusions}   \label{eq:conclusions}

In this paper, we have found the conditions which an observer in Massive gravity has to satisfy in order to agree with the black-hole temperature obtained originally in GR. The conditions impose some constraints in the functional behavior on the spacetime paths which the observers have to follow. The result is general in the sense that it is independent on whether the vacuum solution in massive gravity is single or degenerate, taking into account that massive gravity is a gravitational $\sigma$-model. The~conditions to be satisfied by the St\"uckelberg fields come from the normalization of the Killing vector in the direction of $T_0(r,t)$ defined previously in this paper. {Note that the observers can take any arbitrary path in the spacetime. However, if they follow the paths marked by the constraints over the St\"uckelberg functions, as those defined in this paper, then they will find the same temperature as in GR inside the scenario of massive gravity}. The analysis done in this paper illustrates in addition the coincidence between the scales $r_0$ in GR and $r_V$ in massive gravity. Other authors have derived some interesting analysis about the black-hole thermodynamics in massive gravity \cite{Rgen, Rgen-2}, however~they did not consider the fact that in reality in massive gravity theories, the amount of particles emitted by the event horizon never changes. What in reality is different is the way how the observers define the notions of vacuum in massive gravity. For example, the modification for the black-hole temperatures found in \cite{Rgen, Rgen-2} are correct as far as we consider arbitrary observers located at scales where the extra-degrees of freedom of the theory become relevant and in addition if the observers define the time direction arbitrarily. However, when the gravitational field is strong, the deviations with respect to GR must be small if the theory is consistent. For this reason, in \cite{NR, NR-2}, it was found that the periodicity of the poles of the propagator is distorted for the case of a scalar field moving around a source in massive gravity. This creates an apparent modification of the black-hole temperature in this theory. However, the reality is that if the observer defines the time coordinate in agreement with $T_0(r,t)$, such modifications with respect to GR disappear. {In this paper in addition we have demonstrated that different observers can define different degrees of freedom depending on how they define the time-coordinate with respect to the St\"uckelberg function. An observer defining the local time in agreement with $T_0(r,t)$ will define the same degrees of freedom as in GR. On the other hand, an observer defining the time arbitrarily with respect to $T_0(r,t)$ will perceive more degrees of freedom. This is related to the way how we define the local masses for the different modes $m_{ij}$. These masses will depend on the non-trivial contributions coming from the St\"uckelberg function. By using these approaches, it would be interesting to analyze possible phase transitions for black-hole solutions in a similar way as has been done in \cite{SF}. Other~interesting approaches have been developed in \cite{Hendi, Hendi-2}.                 

\newpage

\appendix

\section{General Solution of a Fourth-Order Polynomial} \label{eq:fos}

The standard form of a fourth-order equation can be taken to be
\begin{equation}   \label{eq:sfoafoeia}
Ax^4+Bx^3+Cx^2+Dx+E=0
\end{equation}
in order to get the reduced form of this equation, we need to do the following variable change
\begin{equation}   \label{eq:cov}
y\equiv x+\frac{B}{4A}.
\end{equation}

Solving for x, we get
\begin{equation}   \label{eq:cov2}
x=y-\frac{B}{4A}.
\end{equation}

Replacing this result inside of Equation~(\ref{eq:sfoafoeia}), we get
\begin{equation}
A\left(y-\frac{B}{4A}\right)^4+B\left(y-\frac{B}{4A}\right)^3+C\left(y-\frac{B}{4A}\right)^2+D\left(y-\frac{B}{4A}\right)+E=0.
\end{equation}

Developing parenthesis, we obtain
\begin{multline}
A\left(y^4-\left(\frac{B}{A}\right)y^3+\frac{3}{8}\left(\frac{B}{A}\right)^2y^2-\frac{1}{16}\left(\frac{B}{A}\right)^3y+\left(\frac{B}{4A}\right)^4\right)+\\
B\left(y^3-\frac{3}{4}\left(\frac{B}{A}\right)y^2+\frac{3}{16}\left(\frac{B}{A}\right)^2y-\frac{1}{64}\left(\frac{B}{A}\right)^3\right)+C\left(y^2-\frac{B}{2A}y+\frac{1}{16}\left(\frac{B}{A}\right)^2\right)\\
+D\left(y-\frac{B}{4A}\right)+E=0.
\end{multline}

Regrouping common factors and dividing by A, we get

\begin{equation}   \label{eq:refor}
y^4+\left(-\frac{3}{8}\left(\frac{B}{A}\right)^2+\frac{C}{A}\right)y^2+\left(\frac{1}{8}\left(\frac{B}{A}\right)^3-\frac{CB}{2A^2}+\frac{D}{A}\right)y-\frac{3}{256}\left(\frac{B}{A}\right)^4+\frac{1}{16}\frac{C}{A^3}B^2-\frac{DB}{4A^2}+\frac{E}{A}=0.
\end{equation}

The standard reduced fourth-order equation is given by
\begin{equation}   \label{eq:reduced}
y^4+Py^2+Qy+K=0.
\end{equation}

Comparing this form with (\ref{eq:refor}), we obtain
\begin{eqnarray}   \label{eq:c}
P=\frac{C}{A}-\frac{3}{8}\left(\frac{B}{A}\right)^2\nonumber && \\
Q=\frac{1}{8}\left(\frac{B}{A}\right)^3-\frac{CB}{2A^2}+\frac{D}{A}\nonumber && \\
K=\frac{E}{A}-\frac{DB}{4A^2}+\frac{1}{16}\frac{CB^2}{A^3}-\frac{3}{256}\left(\frac{B}{A}\right)^4.
\end{eqnarray}

The associated cubic equation of (\ref{eq:reduced}) is
\begin{equation}   \label{eq:t}
z^3+2Pz^2+(P^2-4K)z-Q^2=0,
\end{equation}
where P, Q and K are given in (\ref{eq:c}). Given the solutions of the associated third-order Equation~(\ref{eq:t}), the solutions for the reduced Equation 
(\ref{eq:reduced}) are given by
\begin{eqnarray}   \label{eq:fosaaaa}
y_1=\frac{1}{2}(\sqrt{z_1}+\sqrt{z_2}-\sqrt{z_3})\nonumber && \\
y_2=\frac{1}{2}(\sqrt{z_1}-\sqrt{z_2}+\sqrt{z_3})\nonumber && \\
y_3=\frac{1}{2}(-\sqrt{z_1}+\sqrt{z_2}+\sqrt{z_3})\nonumber && \\
y_4=\frac{1}{2}(-\sqrt{z_1}-\sqrt{z_2}-\sqrt{z_3}).
\end{eqnarray}

The solutions of the associated third-order Equation~(\ref{eq:t}), must satisfy the following
\begin{equation}   \label{eq:ctsa}
Q=\sqrt{z_1}\sqrt{z_2}\sqrt{z_3}.
\end{equation}
 
\section{General Solution of a Third-Order Polynomy} \label{eq:fos2}

The standard form of a third-order polynomy is given by
\begin{equation}   \label{eq:sfoatopaaa}
ax^3+bx^2+cx+d=0.
\end{equation}

The normal form is obtained by dividing with respect to $a$ the result as follows
\begin{equation}   \label{eq:third-order}
x^3+rx^2+sx+t=0,
\end{equation}
where obviously, we have
\begin{equation}   \label{eq:cr}
r=\frac{b}{a}\;\;\;\;s=\frac{c}{a}\;\;\;\;t=\frac{d}{a},
\end{equation}
with the extra condition $a\neq 0$. The reduced form of the third-order Equation~(\ref{eq:third-order}), requires the change of variable
\begin{equation}   \label{eq:covto}
y\equiv x+\frac{r}{3},
\end{equation}
and the reduced form is given by
\begin{equation}   \label{eq:rt}
y^3+py+q=0,
\end{equation}
with the corresponding coefficients given by
\begin{equation}   \label{eq:dc}
p=s-\frac{r^2}{3}\;\;\;\;q=\frac{2}{27}r^3-\frac{rs}{3}+t,
\end{equation}
where $r$, $s$ and $t$ are given in (\ref{eq:cr}). On the other hand, it is necessary to establish a classification criteria for the reduced third-order equation given in (\ref{eq:rt}), the criteria is based in a parameter $D$ given by
\begin{equation}   \label{eq:D}
D\equiv \left(\frac{p}{3}\right)^3+\left(\frac{q}{2}\right)^2.
\end{equation}

Also for the same classification, it is necessary to find an auxiliary parameter given by
\begin{equation}   \label{eq:RR2}
R\equiv sign(q)\sqrt{\frac{\vert p\vert}{3}}.
\end{equation}
where $q$ and $p$ are defined in eqns. (\ref{eq:rt}) and (\ref{eq:dc}). On the other hand, we define an auxiliary angle $\phi$, which is defined depending of the following cases:

\vspace{6pt}
Case (i). $p<0, D\leqslant 0$.
\vspace{6pt}

In this case, the auxiliary angle is defined as
\begin{equation}    \label{eq:auxanfc}
\cos\phi\equiv \frac{q}{2R^3},
\end{equation}
with the corresponding solutions
\begin{eqnarray}   \label{eq:tosfc}
y_1=-2R\cos\frac{\phi}{3}\nonumber && \\
y_2=-2R\cos\left(\frac{\phi}{3}+\frac{2\pi}{3}\right)\nonumber && \\
y_3=-2R\cos\left(\frac{\phi}{3}+\frac{4\pi}{3}\right).
\end{eqnarray}

\vspace{6pt}
Case (ii). $p<0, D>0$.
\vspace{6pt}

In this case, the auxiliary angle is defined as
\begin{equation}   \label{eq:auxansc}
\cosh\phi\equiv \frac{q}{2R^3},
\end{equation}
and the corresponding solutions are
\begin{eqnarray}   \label{eq:tossc}
y_1=-2R\cosh\frac{\phi}{3}\nonumber && \\
y_2=R\cosh\frac{\phi}{3}+i\sqrt{3}R\sinh\frac{\phi}{3}\nonumber && \\
y_3=y_2*=R\cosh\frac{\phi}{3}-i\sqrt{3}R\sinh\frac{\phi}{3}.
\end{eqnarray}

\vspace{6pt}
Case (iii). $p>0, D>0$.
\vspace{6pt}

In this section, we define the auxiliary angle to be
\begin{equation}   \label{eq:auxantc}
\sinh\phi\equiv \frac{q}{2R^3}
\end{equation}
with the corresponding solutions
\begin{eqnarray}   \label{eq:tostc}
y_1=-2R\sinh\frac{\phi}{3}\nonumber && \\
y_2=R\sinh\frac{\phi}{3}+i\sqrt{3}R\cosh\frac{\phi}{3}\nonumber && \\
y_3=y_2*=R\sinh\frac{\phi}{3}-i\sqrt{3}R\cosh\frac{\phi}{3}.
\end{eqnarray}

In the mentioned three cases, $D$ is defined in Equation~(\ref{eq:D}); $p$ is defined in Equations (\ref{eq:rt}) and (\ref{eq:dc}), and the parameter $R$ is defined in Equation~(\ref{eq:RR2}).

\end{document}